\documentclass[aps,prd,preprintnumbers,nofootinbib,twocolumn]{revtex4}
\usepackage{bm}
\usepackage{latexsym}
\usepackage{dcolumn}
\usepackage{amsmath,amsfonts,amssymb}
\usepackage{graphicx,epsfig}
\usepackage{psfrag}
\usepackage{amsthm}

\def\be {\begin{equation}}
\def\ee {\end{equation}}
\def\bea {\begin{eqnarray}}
\def\eea {\end{eqnarray}}
\def\bc {\begin{center}}
\def\ec {\end{center}}
\def\bfg {\begin{figure}}
\def\efg {\end{figure}}
\def\bi {\begin{itemize}}
\def\ei {\end{itemize}}

\def\la {\label}
\def\le {\left}
\def\ri {\right}

\def\no {\noindent}

\def\vs {\vspace}

\def\beq{\begin{equation}}
\def\eeq{\end{equation}}
\def\br{\begin{eqnarray}}
\def\er{\end{eqnarray}}
\newcommand{\eel}[1] {\label{#1}\end{equation}}

\newcommand{\bdm}{\begin{displaymath}}
\newcommand{\edm}{\end{displaymath}}

\begin{document}
\title{Stringent theoretical and experimental bounds on graviton mass
}

\author{Ahmed Farag Ali}
\email{ahmed.ali@fsu.edu.eg}

\affiliation{Department of Physics, Faculty of Science, Benha University, Benha 13518, Egypt}

\author{Saurya Das} \email{saurya.das@uleth.ca}

\affiliation{Theoretical Physics Group, Department of Physics and Astronomy,
University of Lethbridge, 4401 University Drive,
Lethbridge, Alberta, Canada T1K 3M4 \\
{}\\
{\bf This essay received an Honorable Mention in the 2016
Gravity Research Foundation Essay Competition\\
}}

\begin{abstract} %
We show from theoretical considerations, that if the graviton is massive, its mass is constrained to be about $10^{-32}~eV/c^2$.
This estimate is consistent with those obtained from experiments, including
the recent gravitational wave detection in advanced LIGO.
\end{abstract}

\maketitle


%
General Relativity (GR) is one of the most successful physical theories. Its predictions have been verified
over length scales that span over $14$ orders of magnitude, from our solar system to galaxies, galaxy clusters and cosmological scales. An important feature of GR is that gravitational waves predicted from it
(and recently observed \cite{ligo1}) travel at the speed of light, $c = 2.9979 \times 10^8~m/s$. As a result, quanta of gravitational field, or gravitons (assuming that a satisfactory quantum theory of GR would be formulated sometime) should be massless. While this is satisfactory from certain standpoints, and places it at par with the photons, or the quanta of light, we argue in this article that a small but non-zero graviton mass cannot be
ruled out either by theory or by experiments. In fact a small graviton mass may have profound observational and theoretical implications. Furthermore, we show that
tight lower and upper bounds, and hence a reasonable estimate of its mass can be made. In what follows, we first obtain these bounds on the graviton mass from theoretical considerations, and show that these are also consistent with observations. It is therefore important to explore implications of a massive graviton further.

\section{Upper bound on graviton mass}

Exactly how long ranged is gravity? It has been established beyond reasonable doubt that GR is valid
at least in the present epoch,
until the far reaches of our Universe, likely until its edge e.g. as seen from the supernovae 1a data \cite{sn1,sn2},
and also from recent gravitational wave observations, the latter in fact imposing strong constraints on
deviations from GR \cite{yunes}, although not ruling them out completely
\cite{konoplya}.
However, the fact remains that its range has not been, and in fact cannot be tested beyond the cosmological horizon at any epoch, regions outside being causally disconnected from the inside.
This horizon is approximately equal to the Hubble radius $L=c/H$, where $H=$ Hubble parameter.
At present, $H\equiv H_0 \approx 68~km/s/Mpc$ and $L \equiv L_0 \approx 10^{9}$ Mpc. Consequently, from the Yukawa type potential for massive gravity,
\bea
V \propto -\frac{e^{-cmr/\hbar}}{r}~,
\eea
it follows that graviton mass $m \leq \hbar/cL_0~\approx 10^{-32}~eV/c^2$, cannot be ruled out. Any any earlier epoch, $L=L_0a$, where $a=$ the scale factor ($a=1$ in the current epoch, and $\ll1$ in earlier epochs),
and a {\it weaker} bound $m \leq (10^{-32}/a)~eV/c^2$, with $a<1$, would have been obtained.

\section{Lower bound on graviton mass}

It was shown long back that we live in an expanding \cite{hubble} (and in fact accelerating \cite{sn1,sn2}) Universe.
It was also shown rigorously in \cite{higu} that if one attempts to formulate a quantum field theory of a massive gravitational field in an accelerating Universe, then the graviton annihilation and creation operators satisfy the commutator
\footnote{$M$ and $m$ of \cite{higu} are $m$ and $\Lambda/3$ of this article respectively.}
\bea
[b(k), b(k')^\dagger] = \frac{4 (k^2)^2}{3 m^2 (m^2 - \frac{2h^2\Lambda}{3c^2} )} \delta^{(3)} (k-k')~,
\la{higu1}
\eea
where $k,k'$ are the momenta.
Thus in a dark energy dominated Universe at late times, when one can replace $\Lambda \approx 1/L_0^2$,
one concludes from Eq.(\ref{higu1}), that to avoid negative norm states (i.e. ghosts),
equivalently negative probabilities when one introduced interactions,
the graviton mass is bounded from {\it below} by $m \geq \hbar/cL~\approx 10^{-32}~eV/c^2$.
This with the previous upper bound implies that the graviton mass is strongly constrained at $m = 10^{-32}~eV$.
Other theoretical considerations also lead to a graviton mass which is consistent with the above estimate \cite{gravmasstheory1,gravmasstheory2,gravmasstheory3,gravmasstheory3a,gravmasstheory4}.

Once again, in any earlier epoch, this lower bound would have remained the same (since $\Lambda$ is a constant),
and perfectly consistent with the upper bound $m\leq (10^{-32}/a)~eV/c^2$ referred to earlier.

\section{Consistency with observations}

Estimates of the graviton mass have been made from various observations, such as estimates of Hubble radius,
from orbital decay of binary pulsars and other astrophysical observations \cite{gravmassexpt}.
Recently, detection of a transient gravitational wave signal from a black hole merger was reported \cite{ligo1}.
If gravitons were massive, this signal would have reached a time $\Delta t$ seconds {\it after} the arrival of the corresponding light signal of the same event (assuming they were produced at the same time).
This was used in \cite{ligo1} (and using the results of \cite{will})
to examine a bound on the graviton mass $m$
\bea
m < 7 \times 10^{-23} \le( \frac{D}{200} \frac{100}{f} \frac{1}{f\Delta t} \ri)^{-1/2} eV/c^2~.
\la{gravmassbound}
\eea
where the distance from the source $D$ is in Mpc and the (average) frequency of gravitational wave $f$ in Hz.
$f\Delta t$ in the above is bounded from below by $1/\sigma$, where $\rho$ is the signal to noise ratio of the apparatus \cite{will} \footnote{The signal to noise ratio was denoted by $\rho$ in \cite{ligo1,will}.}.
For the detected gravity wave GW150914, using $D=410$ Mpc, $f=100$ Hz and $\sigma=24$, one gets $m<10^{-22}$ eV, which is weaker than the observational upper bound mentioned earlier, but consistent with it.

\section{Implications of a massive graviton}


\subsection{Gravitational wave astronomy}

 A massive graviton would imply that there would potentially be advance notice for the arrival of gravitational waves - light from astrophysical events such as supernovae or black hole mergers would arrive $\Delta t$ seconds {\it before} the gravitational wave from the same event, where
\bea
\Delta t > \frac{2D}{f^2} \le( 10^{21}~m \ri)^2~.
\eea
%
Therefore, for $m = 10^{-32}~eV/c^2$, one gets
\bea
&& \Delta t > \frac{2 \times 10^{-22} D}{f^2}~s~.
\eea
Although the above appears tiny, with continued progress in the measurement of smaller and smaller time intervals,
observation of such a time difference may not only be possible, but
would constitute one of the strongest experimental tests of graviton mass.

\subsection{Source of Dark Matter and Dark Energy}

It was shown in \cite{bhaduri} that a gravitons with mass $m$, being Bosons, can form a Bose-Einstein condensate (BEC), at temperatures less than the critical temperature, given by
\bea
T_c = \frac{\hbar c}{k_B} \le( \frac{\rho\pi^2}{5 m\zeta(3)} \ri)^{1/3}
\la{Tcrit}
\eea
Furthermore, identifying the density of the condensate, $\rho$ with the dark matter density of our universe, i.e. $\rho=0.25\rho_{crit}/a^3$, where
$\rho_{crit}=10^{-26} kg/m^3$ is the current critical density of our Universe and $m$ expressed in $eV/c^2$, Eq.(\ref{Tcrit}) becomes \cite{bhaduri,dasinflation}
\bea
T_c = \frac{3}{m^{1/3} a}~K
\la{Tcrit2}
\eea
From Eq.(\ref{Tcrit2}) it follows that for $m\leq 1~eV/c^2$, that the background temperature of our Universe
at any epoch, $T(a) = 2.7/a < T_c (a)$. That is the critical temperature exceeds the background temperature of the Universe $T(a)$ at all times, and a giant BEC consisting of gravitons in their ground state
would have formed at a very early epoch.
This BEC may thus account for the dark matter content in our Universe (with its density variations accounting for the dark matter density profiles). Furthermore, under the assumption of large scale homogeneity and isotropy, the quantum potential associated with this wavefunction can also account for its dark energy content \cite{dascoincidence,alidas}. If true, this would provide an unifying picture of dark matter and dark energy. This would also resolve the initial (Big Bang) singularity, since it can be shown that the `quantal trajectories' of points within this BEC, when extrapolated back in time, come very close, yet do not meet at a point.
This follows from the well-known no-crossing property of these trajectories \cite{qre,alidas}
\footnote{There has been some criticism of the quantum potential and its application to the Quantum Raychaudhuri Equation, which is related to some of the results of this section \cite{flawed-Lashin}. As pointed out in \cite{Ali:2015xma}, the authors of \cite{flawed-Lashin} start with incorrect assumptions, and arrive at equally flawed and irrelevant conclusions. Regardless of their incorrect expressions for the quantum corrections, as explained in \cite{Ali:2015xma,qre},
the absence of conjugate points and the resultant invalidity of the singularity theorems in a {\it fixed} classical background,
the everlasting nature of these trajectories, and the no-requirement of quantal trajectories to remain timelike throughout, simply follow from the first order nature of the guiding equations, and the equivalence between quantum wave equations and the equations for quantal trajectories. Furthermore the BEC critical temperature (Eqs.(\ref{Tcrit}-\ref{Tcrit2})) depend on the graviton mass alone.
None of the above depend on the precise form of the quantum corrections.}.
In other words, a small graviton mass may be able to address all of the above issues at one stroke.

\section{Conclusions}

We showed in this essay that not only is a tiny graviton mass allowed by theory as well as experiments, but that if non-zero, then it is tightly constrained to be about $10^{-32}~eV/c^2$, no more, no less. While it is still possible that graviton mass is strictly zero (like that of the photon), a non-zero graviton mass would have some distinct advantages, such as forming a cosmic BEC accounting for dark matter and dark energy, and assisting in the resolution of the initial cosmological singularity. Furthermore, it would also open up the possibility of observing astrophysical events sequentially, first via light and then via gravitational waves emitted from them.
Such an {\it advance warning} for observing weak gravitational waves 
would clearly be useful.
Note that graviton with above mass would mean gravity would be rendered `short ranged' in the future, when
the cosmological horizon $L\gg L_0$, and deviations from GR should result.
In this way, the current epoch appears to play a special role, simply because it marks the era in which the cosmological constant or dark energy starts dominating the matter content of our Universe.
Thus there is all the more reason for studying various implications of a small graviton mass, both from the theoretical as well as the observational side. Although there has been some progress in incorporating graviton mass in a covariant theory of gravity (i.e. extension of general relativity) \cite{gravmasstheory3,gravmasstheory5,gravmasstheory6},
much more needs to be done to have a satisfactory theory.
We hope that there will be progress in the above directions in the near future.

\vs{.2cm}
\no {\bf Acknowledgment}

\no
This work is supported by the Natural Sciences and Engineering
Research Council of Canada.



\end{document}